\documentclass[12pt]{iopart}
\newcommand{\la}{\langle}
\newcommand{\ra}{\rangle}
\newcommand{\up}{\uparrow}
\newcommand{\dn}{\downarrow}
\newcommand{\vk}{{\bf k}}

\newcommand{\vp}{{\bf p}}

\newcommand{\vx}{{\bf x}}
\newcommand{\vy}{{\bf y}}
\newcommand{\vr}{{\bf r}}
\newcommand{\vQ}{{\bf Q}}
\newcommand{\vA}{{\bf A}}
\newcommand{\be}{\begin{equation}}
\newcommand{\ee}{\end{equation}}
\newcommand{\bea}{\begin{eqnarray}}
\newcommand{\eea}{\end{eqnarray}}

\usepackage{iopams, graphicx}  
\begin{document}

\title{Single vortex structure in two models of iron pnictide $s^\pm$ superconductivity}

\author{M. A. N. Ara\'ujo$^{1,2}$, M. Cardoso$^3$ and P. D. Sacramento$^1$}

\address{$^1$CFIF, Instituto Superior T\'ecnico, 
Universidade T\'ecnica de Lisboa, Av. Rovisco Pais, 1049-001 Lisboa, Portugal }
\address{$^2$ Departamento de F\'{\i}sica,  Universidade de \'Evora, P-7000-671, \'Evora, Portugal}
\address{$^3$CFTP, Instituto Superior T\'ecnico,
Universidade T\'ecnica de Lisboa, Av. Rovisco Pais, 1049-001 Lisboa, Portugal }
\ead{mana@evunix.uevora.pt}
\begin{abstract}
The structure of a single vortex in a FeAs superconductor is studied, in the framework of 
two formulations of superconductivity for the recently proposed 
sign-reversed $s$ wave ($s^\pm$) scenario: {\it (i)} a continuum model      
taking into account the existence of an electron and a hole band with a repulsive local interaction 
between the two; {\it (ii)} a lattice tight-binding model with two orbitals per unit cell 
and a next-nearest-neighbour attractive interaction.  
In the first model, the  local density of states (LDOS) at the vortex centre, as a function of energy, 
exhibits a peak at the Fermi level,
while in the second model such LDOS peak is deviated from the Fermi level and its energy depends on band filling. 
An impurity located outside the vortex core has little effect on the LDOS peak, but an impurity close to the vortex core
can almost suppress it and modify its position. 
\end{abstract}
\maketitle

\section{Introduction}

 The discovery of superconductivity in iron based materials at relatively high temperatures 
quickly  spurred intensive  theoretical and experimental work \cite{norman}.
The electron-phonon interaction has early been ruled out as the pairing mechanism \cite{boeri} 
and spin fluctuation mediated interactions have been proposed \cite{mazin,tesanovic,scalapino}. 
The most studied materials can be divided in three families: the  "lanthanum" family LaFeAsO$_{1-x}$F$_x$, 
the "rare-earth" family (Sm, Nd, Ce, Pr)FeAsO$_{1-x}$F$_x$, and the (Ba, Ca, Sr)Fe$_2$As$_2$ family
(also known as "122" compounds).   
 
Iron pnictide superconductors have a  complex band structure,
with a Fermi surface (FS) consisting of four sheets,
two of them hole-like, and the other two electron-like \cite{mazin,singh,xu,haule}.
S-wave, d-wave and p-wave pairing scenarios have early been proposed to describe the
superconducting state \cite{chubukovs,qi,lee,yao}, as well as extended s-wave
\cite{graser}. More recently, two
possible pairing scenarios have been in dispute:
 the "sign-reversed s-wave state" (s$^\pm$-state), where the gap function
has no nodes on any of the FS sheets but takes on
opposite signs on the hole-like and electron-like sheets of the FS \cite{mazin,tesanovic}; 
the extended s-wave symmetry, in which
the gap function has four nodes on the electron FS sheets, no nodes on the hole sheets,
and has the same sign on both hole FS's \cite{scalapino}. 
The existing experimental evidence for lanthanum based superconductors
seems to be mostly consistent with unconventional gap functions with nodes on the FS, 
while the experiments with superconductors containing  rare-earth atoms 
as well as with the 122 compounds tend to suggest a nodeless nearly
 isotropic gap across the FS. However,
Andreev reflection studies have lead
to different conclusions regarding  the  pairing symmetry for
the same (rare-earth based) material \cite{tesanovicandreev,chineses}.

The mixed phase of the iron pnictide superconductors has been less studied.  
Specific heat measurements in LaFeAsO$_{1-x}$F$_x$ under a magnetic field revealed a 
$\sqrt{H}$ dependence of the specific heat coefficient,  
which has been interpreted as due to the Doppler shift of low energy excitations
(close to the nodes of the gap function) caused by the 
superfluid flow far from  the vortex cores \cite{mu,scalapinovortex, volovik}.  
More recently, a scanning tunnelling microscope (STM) observation of vortices in 
BaFe$_{1.8}$Co$_{0.2}$As$_2$ has been made \cite{experiment}, where 
the differential conductance versus bias voltage was measured  inside and 
outside the vortex cores. 
On the sample surface scanned by the STM, the gap value (6 meV) was almost uniform 
across the surface, with a fractional variance of 12\%, and 
the vortex locations were not coincident with the impurities present in the material. 
The observed differential conductance  inside the vortex
cores  displayed no peaks signalling the presence of core bound states, 
only the V-shaped background.      
   
The  s$^\pm$  superconducting state, because it is novel,  
has been the subject of intense theoretical investigation. Such studies
often predict unconventional  behaviour of various physical observables, 
qualitatively  different from the usual s-wave state. This is because the
gap function, despite having no nodes on the FS sheets, changes sign between them, 
effectively leading to a Josephson-like coupling between two s-wave gap functions 
with a phase difference of $\pi$, and also because of quantum interference effects 
between  Bloch waves belonging to  different FS pockets.   

In this work we study the single vortex structure in the s$^\pm$  state, using two different
formulations.  
In the continuum  formulation, only two FS sheets are considered, 
having parabolic dispersions with opposite masses.   
In the discrete formulation, a tight-binding model that reproduces the 
band structure close to the  Fermi level is used. In this model,   
we find the lowest energy (subgap) excitations to be spatially extended, in contrast to the
case with conventional s-wave vortices and the results of the continuum formulation, 
where the excitations are localized near the 
vortex core. Only higher energy excitations (but still subgap) are localized.  
This effect is a manifestation of interband interference effects which have earlier
been found to produce Andreev bound states at finite energy \cite{us}.       
We also study the effect of impurities, close to the vortex core, on the local density of states 
and find that they  have a strong effect on local density of states at the vortex core.

\section{Continuum formulation}

\subsection{Model}

We begin by considering a simplified model for the FeAs as described
in ref. \cite{bang}. The Hamiltonian is of the form
\bea
\hat H &=& \sum_{\vk,\sigma} \epsilon_h(\vk) \hat h_{\vk,\sigma}^{\dagger}  \hat h_{\vk,\sigma}+
 \sum_{\vk,\sigma} \epsilon_e(\vk) \hat e_{\vk,\sigma}^{\dagger}  \hat e_{\vk,\sigma} \nonumber \\
&+& \sum_{\vk,\vk'} V_{hh}(\vk,\vk') \hat h_{\vk,\uparrow}^{\dagger} \hat h_{-\vk,\downarrow}^{\dagger}
\hat h_{-\vk',\downarrow} \hat h_{\vk',\uparrow} \nonumber \\
&+& \sum_{\vk,\vk'} V_{ee}(\vk,\vk') \hat e_{\vk,\uparrow}^{\dagger} \hat e_{-\vk,\downarrow}^{\dagger}
\hat e_{-\vk',\downarrow} \hat e_{\vk',\uparrow} \nonumber \\
&+& \sum_{\vk,\vk'} V_{he}(\vk,\vk') \hat h_{\vk,\uparrow}^{\dagger} \hat h_{-\vk,\downarrow}^{\dagger}
\hat e_{-\vk',\downarrow}  \hat e_{\vk',\uparrow} \nonumber \\
&+& \sum_{\vk,\vk'} V_{eh}(\vk,\vk') \hat e_{\vk,\uparrow}^{\dagger} \hat e_{-\vk,\downarrow}^{\dagger}
\hat h_{-\vk',\downarrow} \hat h_{\vk',\uparrow}
\eea
The model describes two bands, one of positive mass particles (electrons), $\hat e_{\vk,\sigma}$,  with dispersion
$\epsilon_e(\vk) $ and another of negative mass particles (holes), $\hat h_{\vk,\sigma}$,  with dispersion $ \epsilon_h(\vk)$ 
in two dimensions.  
The hole band  is centred at the origin of the Brillouin zone ($\Gamma$ point),
 while the electron band 
 is centred at the M point, $(\pi, \pi)$. 
It is then assumed
that the two types of particles interact via pair intraband couplings ($V_{hh},V_{ee}$)
and pair interband (or Josephson)  couplings ($V_{he},V_{eh}$). These interactions are supposed to
have a magnetic origin and are taken as repulsive \cite{bang,korshunov}.
In order to do  a continuum formulation of the vortex problem in real space, we take the $h$ band 
as  parabolic with negative mass and the $e$  band as a parabola of positive
mass. The bottom of the $e$ band is located at $E_{0}^{e} = -0.58$, the top of the $h$ band
is located at $E_{0}^{h} = 0.24$ (the energies are in eV units). 
The chemical potential is located at $E_F=0$. The masses of the two bands
are taken as $m_e=1, m_h=-1.8$. 
Since the $h$ particles have negative mass, 
it is convenient to make a particle-hole transformation  
as $\hat h_{\sigma}(\vr) \rightarrow \tilde{h}_{-\sigma}^{\dagger}(\vr)$.
We also  perform the gauge transformation
$\bar{e}_{\sigma}(\vr)\rightarrow e^{i \vQ \cdot \vr} \hat e_{\sigma}(\vr)$, with $\vQ=(\pi, \pi)$,  
on the electrons. 
A standard mean field decoupling of the interaction terms leads to the Bogolubov-de
Gennes (BdG) equations for the wave functions of the quasiparticles \cite{deGennesbook}
\bea
\left[ \frac{\hat\vp^2}{2 m_e} + E_{0}^{e} - E_F \right] \bar{u}_e(\vr) + \Delta_e(\vr) \bar{v}_e(\vr) &=& E^e
\bar{u}_e(\vr) \nonumber \\
 \Delta_e^*(\vr) \bar{u}_e(\vr) -\left[ \frac{\hat \vp^2}{2 m_e} + E_{0}^{e} -E_F \right] \bar{v}_e(\vr)  &=& E^e
\bar{v}_e(\vr) \nonumber \\
\left[ \frac{\hat \vp^2}{2 |m_h|} - E_{0}^{h} + E_F \right] \tilde{u}_h(\vr) + \Delta_h(\vr) \tilde{v}_h(\vr) &=&
E^h \tilde{u}_h(\vr) \nonumber \\
\Delta_h^*(\vr) \tilde{u}_h(\vr) -\left[ \frac{\hat \vp^2}{2 |m_h|} - E_{0}^{h} + E_F \right] \tilde{v}_h(\vr)  &=& 
E^h \tilde{v}_h(\vr)\,,
\eea
where $\hat \vp=-i\hbar\nabla$ in two dimensions. 
Here, $\bar{u}_e(\vr),\bar{v}_e(\vr)$ are the usual BdG components of the wave function for
the electrons $\hat e$ and  
$\tilde{u}_h(\vr),\tilde{v}_h(\vr)$ are the wave function components for the holes $\tilde{h}$  
(which now have positive mass). 
The self-consistency
of the mean-field approach implies that
\bea
\Delta_e(\vr) &=& -V_1 \sum_{0 \leq E_n^e \leq \hbar \omega_D} \bar{u}_{e n}(\vr)
\bar{v}_{e n}^*(\vr) \left[ 1- 2 f(E_n^e) \right] \nonumber \\
&-& V_2 \sum_{0 \leq E_n^h \leq \hbar \omega_D} \tilde{u}_{h n}^*(\vr)
\tilde{v}_{h n}(\vr) \left[ 1- 2 f(E_n^h) \right] \nonumber \\
\Delta_h(\vr) &=& -V_1 \sum_{0 \leq E_n^h \leq \hbar \omega_D} \tilde{u}_{h n}(\vr)
\tilde{v}_{h n}^*(\vr) 
\left[ 1- 2 f(E_n^h) \right] \nonumber \\
&-& V_2 \sum_{0 \leq E_n^e \leq \hbar \omega_D} \bar{u}_{e n}^*(\vr)
\bar{v}_{e n}(\vr) \left[ 1- 2 f(E_n^e) \right]
\eea
where $f(E)$ is the Fermi function. We have approximated the pair couplings
by two constants $V_1=V_{hh}=V_{ee}$ and $V_2=V_{he}=V_{eh}$. 
We may consider the insertion of a magnetic field in the usual minimal coupling way
taking $\vp \rightarrow \vp-\frac{q}{c} \vA$, where $\vA$ is the
vector potential. The charge $q$ is negative for $\hat e$ electrons ($q=e<0$) and 
positive for the  $\tilde{h}$ holes  ($q=-e$).   
The vector potential is given by Maxwell's equations
\begin{equation}
\nabla \times \mathbf{B} = \nabla \times \nabla \times \mathbf{A}
 = \frac{4 \pi}{c}\mathbf{J}_{total}
\end{equation}
which, in the Coulomb gauge ( $\nabla . \mathbf{A} = 0$ ), is given by
\begin{equation}
\nabla^2 \mathbf{A} = - \frac{4\pi}{c} \mathbf{J}_{total}
\end{equation}
The current density originated in the supercurrents is obtained self-consistently by
\bea
\mathbf{J}(\mathbf{r}) &=& \frac{e\hbar}{im_e} \sum_n 
\left\{
f(E_n) \bar{u}_{e n}^{*}(\mathbf{r})
\left[ \nabla - \frac{ie}{\hbar c}\mathbf{A}(\mathbf{r}) \right] \bar{u}_{e n}
(\mathbf{r})\right.  \nonumber \\ 
&+& \left.  [(1-f(E_n)] \bar{v}_{e n}(\mathbf{r}) \left[ \nabla -
\frac{ie}{\hbar c}\mathbf{A}(\mathbf{r})\right] \bar{v}_{e n}^{*}(\mathbf{r})
\right\} - c.c. \nonumber \\
&-& \frac{e\hbar}{i |m_h|} \sum_n \left\{ 
f(E_n) 
\tilde{u}_{h n}^{*}(\mathbf{r})
\left[ \nabla + \frac{ie}{\hbar c}\mathbf{A}(\mathbf{r}) \right] \tilde{u}_{h n}
(\mathbf{r}) \right. \nonumber \\
&+& \left. [1-f(E_n)] \tilde{v}_{h n}(\mathbf{r}) \left[ \nabla + 
\frac{ie}{\hbar c}\mathbf{A}(\mathbf{r})\right] \tilde{v}_{h n}^{*}(\mathbf{r})\right\} - c.c.
\eea

We write the order parameters using polar coordinates in the form
\bea
\Delta_e(\mathbf{r}) &=& \Delta_e(\rho) e^{-i \varphi} \nonumber \\ 
\Delta_h(\mathbf{r}) &=& \Delta_h(\rho) e^{i \varphi}  
\eea
where $\rho$ and $\varphi$ and the radial and angular polar coordinates. 
Such a two-component vortex  has magnetic flux equal to one flux quantum
( $\Phi= \Phi_0 = \frac{hc}{2|e|}$ ): this is because the diamagnetic currents
from electrons and holes add up, screening the external magnetic field
to  one flux quantum \cite{babaev}.
The system is placed in a cylinder of radius $R$.
Given the azimuthal symmetry of the system, neither $\Delta(\rho)$
nor $\mathbf{A}$ depend on $\varphi$. Therefore the Hamiltonian
may be diagonalized separately for each value of the angular
momentum, $\mu$, which becomes a good quantum number. 
Following  Ref. \cite{Gygi}, the wave functions $\bar{u}_n$ and $\tilde{v}_n$ 
are written in the form 
\begin{equation}
\bar{u}_{e n}(\mathbf{r}) = \bar{u}_{e n}(\rho) e^{i(\mu-1/2)\varphi}
\end{equation}
\begin{equation}
\bar{v}_{e n}(\mathbf{r}) = \bar{v}_{e n}(\rho) e^{i(\mu+1/2)\varphi}
\end{equation}
\begin{equation}
\tilde{u}_{h n}(\mathbf{r}) = \tilde{u}_{h n}(\rho) e^{i(\mu+1/2)\varphi}
\end{equation}
\begin{equation}
\tilde{v}_{h n}(\mathbf{r}) = \tilde{v}_{h n}(\rho) e^{i(\mu-1/2)\varphi}
\end{equation}
where  the radial functions are expanded in a way similar to ref. \cite{Gygi,cardoso}
using as basis functions
\begin{equation}
\phi_{jm}(\rho) = \frac{\sqrt{2}}{RJ_{m+1}(\alpha_{jm})}J_{m}
\left(\alpha_{jm}\frac{\rho}{R}\right)
\end{equation}
The functions $J_m$ are the cylindrical Bessel functions and
$\alpha_{jm}$ is the $j^{th}$ zero of the Bessel function of order $m$.
The set of values of the angular momentum is given by $\mu=\pm (2l+1)/2$ where
$l=0,1,2,\cdots$.

To illustrate the procedure we consider the standard case of the e bands.
For each eigenvalue $E_n$, we have a single value of $\mu$ and it is enough
to diagonalize the matrix, defined in the subspace of the zeros of the Bessel
function,
\begin{equation}
\left( \begin{array}{cccc}
                  T^{-} & \Delta \\
                  \Delta^T & T^{+} \\
                  \end{array}
                  \right)
\left( \begin{array}{c}
                  c_{\mu}^n  \\
                  d_{\mu}^n \\
                  \end{array}
                  \right)
= E_n
\left( \begin{array}{c}
                  c_{\mu}^n  \\
                  d_{\mu}^n \\
                  \end{array}
                  \right)
\end{equation}
where
\bea
T_{j j'}^{-} &=&  \frac{\hbar^2}{2m} \frac{\alpha_{j,\mu-1/2}^2}{R^2}
\delta_{jj'} - (\mu - 1/2) \frac{e}{\hbar c} I_1^- 
+ \frac{e^2}{\hbar^2 c^2}I_2^- + ( E_{0}^{e} - E_F ) \delta_{jj'}
\eea
\bea
T_{j j'}^{+} &=& - \frac{\hbar^2}{2m} \frac{\alpha_{j,\mu-1/2+n}^2}{R^2}
\delta_{jj'} - (\mu-1/2+n) \frac{e}{\hbar c} I_1^+ 
- \frac{e^2}{\hbar^2 c^2}I_2^+ + ( E_F - E_{0}^{e} ) \delta_{jj'} \nonumber \\
& &
\eea
with
\begin{equation}
I_1^- = \int_{0}^{R} \phi_{j,\mu - 1/2}(\rho) \frac{A_\varphi(\rho)}
{\rho} \phi_{j',\mu - 1/2}(\rho) \rho d\rho
\end{equation}
\begin{equation}
I_1^+ = \int_{0}^{R} \phi_{j,\mu-1/2+n}(\rho) \frac{A_\varphi(\rho)}
{\rho} \phi_{j',\mu-1/2+n}(\rho) \rho d\rho
\end{equation}
and
\begin{equation}
I_2^- = \int_{0}^{R} \phi_{j,\mu-1/2}(\rho) A_\varphi(\rho)^2
\phi_{j',\mu- 1/2}(\rho) \rho d\rho
\end{equation}
\begin{equation}
I_2^+ = \int_{0}^{R} \phi_{j,\mu -1/2+n}(\rho) A_\varphi(\rho)^2
\phi_{j',\mu - 1/2+n}(\rho) \rho d\rho
\end{equation}
Also we have
\begin{equation}
\Delta_{jj'} = \int_{0}^{R} \phi_{j,\mu - 1/2}(\rho)
\Delta(\rho) \phi_{j',\mu - 1/2+n}(\rho) \rho d\rho
\end{equation}
In the case of the $h$ band the angular momenta are interchanged.
It is important to note that the symmetry of the BdG equations
$u_n(\mathbf{r}) \rightarrow v_n^{*}(\mathbf{r}),
v_n(\mathbf{r}) \rightarrow -u_n^{*}(\mathbf{r}),
E_n \rightarrow - E_n$
allows to reduce the solution to the positive values of $\mu$.
We obtain the eigenvectors and eigenvalues for negative values of $\mu$ using the
above symmetry.
The vector potential is also solved self-consistently, using Poisson's equation,
as explained in \cite{cardoso}.
From the self-consistent solution of the BdG equations we obtain
 the energy spectrum, the magnetic flux, the current, the magnetic
field and the order parameter profiles as a function of distance from the
vortex core. 
In general the contribution of the vector potential in the BdG equations is
negligible but not in the calculation of the supercurrents. Also, to obtain
the exponential decay of the magnetic field we have to take into account its
effect carefully \cite{Gygi,cardoso}. 
The results are shown below.

We also compute the local density of states  (LDOS) at site $\vr$ and energy $E$, defined as
\begin{equation}
\rho_e(E, \vr) \propto - \sum_n \left[  |u_n(\vr)|^2 f'(E-E_n) 
+  |v_n(\vr)|^2 f'(E+E_n)\ \right] \,,  
\label{sempre}
\end{equation}
for the  $e$ electrons, and
\begin{equation}
\rho_{\tilde h}(E, \vr) \propto - \sum_n \left[  |u_n(\vr)|^2 f'(E+E_n) +  |v_n(\vr)|^2 
f'(E-E_n)\ \right] \,,  
\end{equation}
for the holes. The local density of states is observable through STM (for a recent review
see ref. \cite{fischer}).

\subsection{Results}

\begin{figure}[t]
\vspace{1.3cm}
\centerline{\includegraphics[width=6.5cm]{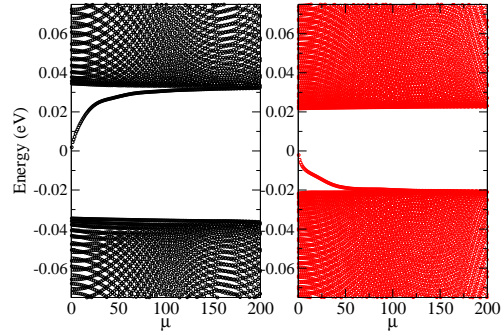}}
\caption{\label{energias} Energy eigenvalues as a function of the angular
momentum for the electron and hole bands.
}
\end{figure}

\begin{figure}[htb]
\vspace{0.3cm}
\centerline{\includegraphics[width=6.5cm]{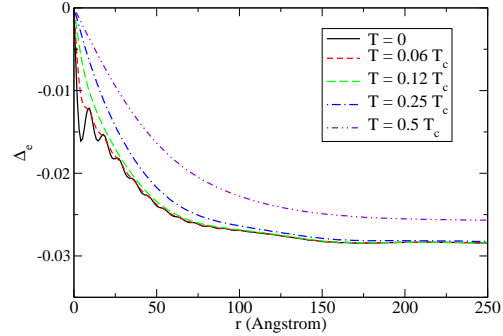}}
\caption{\label{deltae} Order parameter for the electron band at several temperatures.
}
\end{figure}

\begin{figure}[htb]
\vspace{0.7cm}
\centerline{\includegraphics[width=6.5cm]{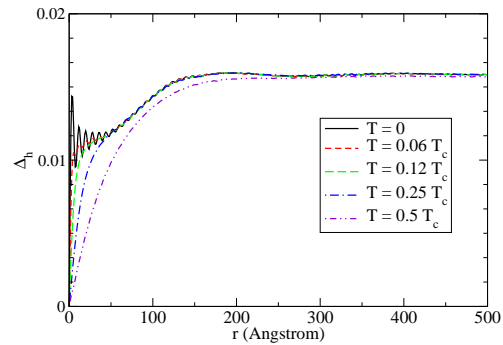}}
\caption{\label{deltah} Order parameter for the hole band.
}
\end{figure}

\begin{figure}[t]
\vspace{0.7cm}
\centerline{\includegraphics[width=6.5cm]{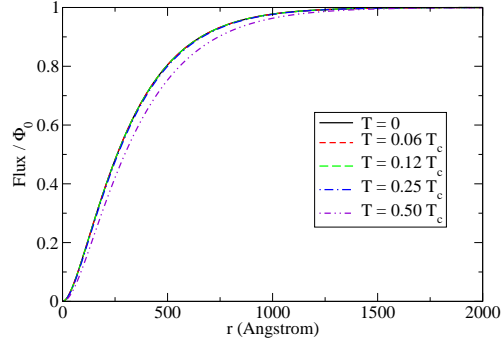}}
\caption{\label{fluxo} Magnetic flux as a function of distance from the vortex core.
}
\end{figure}

\begin{figure}[htb]
\vspace{0.7cm}
\centerline{\includegraphics[width=6.5cm]{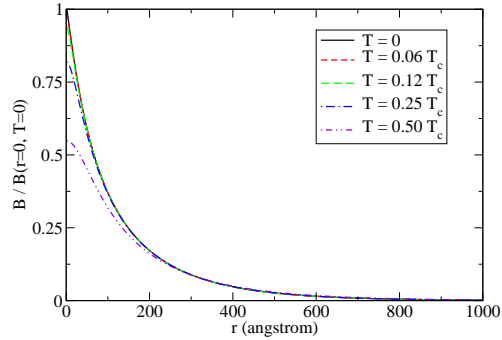}}
\caption{\label{bfield} Magnetic field as a function of distance from the vortex core.
}
\end{figure}

\begin{figure}[htb]
\vspace{0.7cm}
\centerline{\includegraphics[width=6.5cm]{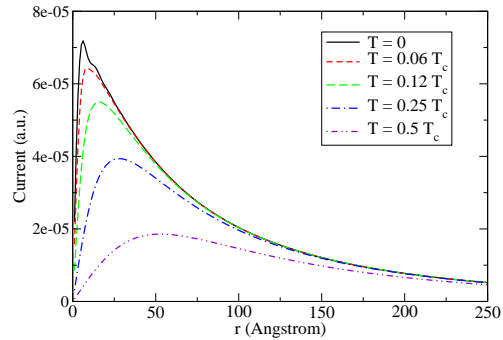}}
\caption{\label{corrente} Total current as a function of distance from the vortex core.
}
\end{figure}

In Figure \ref{energias} we show the energy spectrum for the two bands
for a characteristic set of parameters. We take $V_1=0.6$ and $V_2=0.07$.
In the case of the electron ($e$) band there is a bound state at positive
energies for each value of the angular momentum. These are the well known
Caroli-de Gennes-Matricon states \cite{deGennesbook}. The other states are extended. 
In the case of the hole band ($h$) band the bound states appear at
negative energies (or at positive energies for negative values of the angular momentum).
The spectrum is not symmetric around the Fermi level, as expected \cite{Gygi}. The
bound states branch is symmetric with respect to the sign of the angular momentum.

In Figs. \ref{deltae}, \ref{deltah} we show the spatial dependence of the radial
part of the gap functions for the electron and the hole bands.
As shown before, the two gap functions have opposite signs, characteristic
of the sign-reversed s-wave scenario. At short distances and low temperatures
the gap functions oscillate due to the states inside the core \cite{machida}. As the temperature
grows the oscillations decrease in size, and the size of the core increases
(Kramer-Pesch effect \cite{kramer}). The highest temperature considered is of the order
of half the critical temperature. 

In Fig. \ref{fluxo} we show the magnetic flux enclosed by the vortex as a function
of the distance to the vortex core, saturating to a quantum of flux, as determined
by the choice of the vorticity of the gap functions \cite{cardoso,berthod,monopoles}.
Fig. \ref{bfield} shows the exponential decay of the absolute value of the
magnetic field. 

The current is the result of the electron and hole contributions.
This is shown in Fig. \ref{corrente}. The superfluid current shields the
external field that penetrates through the vortex. Both
contributions to the supercurrent have the same (positive) sign.

Since the gap function of the holes is smaller, the oscillations at small temperatures
are larger. Also, since the energy scale where the gap function tends to the bulk
value is smaller, the spatial length scale is larger in the sense that the fluctuations
extend to larger distances. 
This also affects the oscillations of the current,
which occur, at low temperatures, at two length scales, one associated
with the $e$ band and the other with the $h$ band.

\begin{figure}[htb]
\vspace{0.8cm}
\centerline{\includegraphics[width=8.0cm]{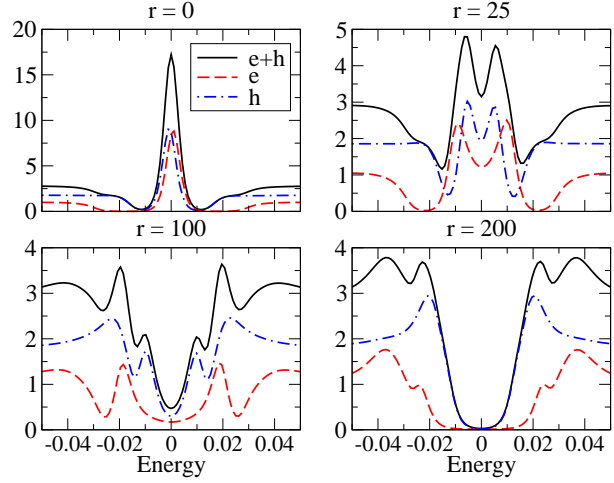}}
\caption{\label{Ldos} LDOS as a function of energy for various
distances from the vortex core. At the vortex there is a zero bias peak.
Moving away from the vortex core, 
a double peak structure of the total LDOS emerges due to the different gaps of
the $h$ and $e$ bands. 
}
\end{figure}

In Fig. \ref{Ldos} we show the LDOS at a set of points whose distance from the vortex
core increases. First we consider a point at the vortex core (simulating the location
of the tip of the STM on top of the vortex core location). We see that the contributions
from the electrons and the holes are not symmetric around the zero bias, reflecting
the existence of bound states at positive and negative energies, respectively.
However, the total LDOS resulting from the sum of the two contributions is quite
symmetric. We find, therefore, that in this simple model there is a zero bias
peak at the vortex core. As we move further from the vortex core position, the central
peak splits, leading to two contributions that clearly separate. This is due to the
"coherence" peaks which reflect the size of the gap function as we move away from the
vortex core and approximate the LDOS of the bulk. The LDOS has a gap which is due
to the smaller of the gaps (in this case originates in the $h$ band) and shows
a double peak structure due to the different $h$ and $e$ band gaps.

\section{Discrete formulation}

\subsection{Model}
A recent tight-binding model \cite{raghu} for the band structure of iron pnictide 
superconductors assumes 
two orbitals per unit  cell, $\rm d_{xz}$ and $\rm d_{yz}$, in a square lattice 
(we take the lattice constant equal to unity).
The tight-binding Hamiltonian in real space can be written as 
\begin{eqnarray}
\hat H &=& \sum_{\nu=x,y}\sum_{i,j,\sigma} \left[ \left( -t_{ij\sigma, \nu} -E_F\delta_{ij}\right) 
\hat d^\dagger_{i\sigma, \nu} \hat d_{j\sigma, \nu}
-t_4 \Theta_{ij} \hat d^\dagger_{i\sigma, \nu} \hat d_{j\sigma, \bar{\nu}} \right] \nonumber\\
&+& \frac{1}{4} \sum_{\nu=x,y}\sum_{i,j} \Delta_{ij} \hat d^\dagger_{ij\up, \nu} \hat d^\dagger_{ij\dn, \nu}
+ {\rm H.c.} \,,
\label{raghu}
\end{eqnarray}
where $\sigma=\up, \dn$ denotes spin projections, the indices  $i,j$ denote lattice sites,
$\nu=x,y$ denotes the atomic orbitals  $\rm d_{xz}$ and $\rm d_{yz}$ at a lattice site, and $\{ \nu, \bar \nu\}=\{ x, y \}$. 
The first neighbour hoppings along the $x$ direction are $t_{ij\sigma, x}=t_1$,  $t_{ij\sigma, y}=t_2$; 
first neighbour hoppings along the $y$ direction are $t_{ij\sigma, x}=t_2$,  $t_{ij\sigma, y}=t_1$. The 
second neighbour hoppings are  $t_{ij\sigma, \nu}=t_3$ and $t_4$.
The  hopping amplitude $t_4$ in equation (\ref{raghu}) appears multiplied by  
$\Theta_{ij}=1$ if the vector connecting the second neighbours $\vec{ij}=\pm (1,-1)$
and $\Theta_{ij}=-1$ if  $\vec{ij}=\pm (1,1)$. 
Note that $\Theta_{ij}$ breaks the fourfold symmetry explicitly.  
The model parameters are $t_1=-1$, $t_2=1.3$, $t_3=t_4=-0.85$. The choice $E_F=1.45$ produces 
the four Fermi surface pockets in the downfolded Brillouin zone (two electron and two hole sheets),
as observed in photo-emission  experiments and {\it ab initio} calculations \cite{raghu}. 
It has been pointed out that the 
 Fermi velocity in the electron Fermi surface sheets becomes underestimated 
in model (\ref{raghu}), however \cite{scalapino}.
\begin{figure}[htb]
\centerline{\includegraphics[width=6.5cm]{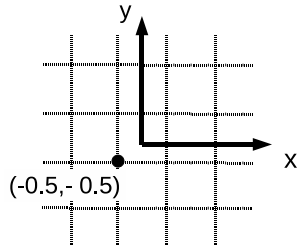}}
\caption{\label{esquema} The two-dimensional square lattice used in the calculations. 
The vortex centre has coordinates $(0,0)$, in the centre of a plaquette.  
The point $(-0.5, -0.5)$, where the LDOS is computed below, is shown.} 
\end{figure}
The above Hamiltonian contains superconductivity at   the mean field level through the 
gap function $\Delta_{ij}$. The choice  of a constant $\Delta_{ij}=\Delta$,
with $i$ and $j$ second-neighbours,
 would produce the gap function in momentum space $\Delta(\vk)= \Delta\cos(k_x)\cos(k_y)$,
corresponding to the proposed sign-reversed $s$ wave ($s^\pm$) 
scenario of superconductivity \cite{mazin,tesanovic}.
\begin{figure}[t]
\centerline{\includegraphics[width=6.5cm]{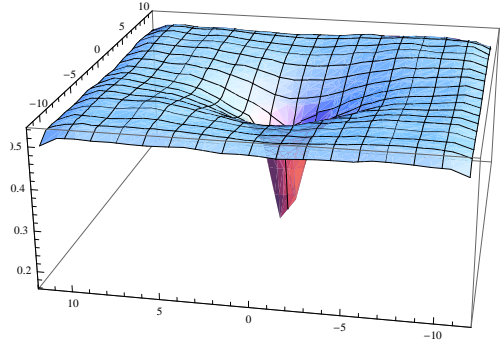}}
\caption{\label{views} Spatial variation of the vortex $s$ wave component, $\Delta_s(i)$, near the vortex core. 
$E_F=1.6$.}
\end{figure}
\begin{figure}
\centerline{\includegraphics[width=6.5cm]{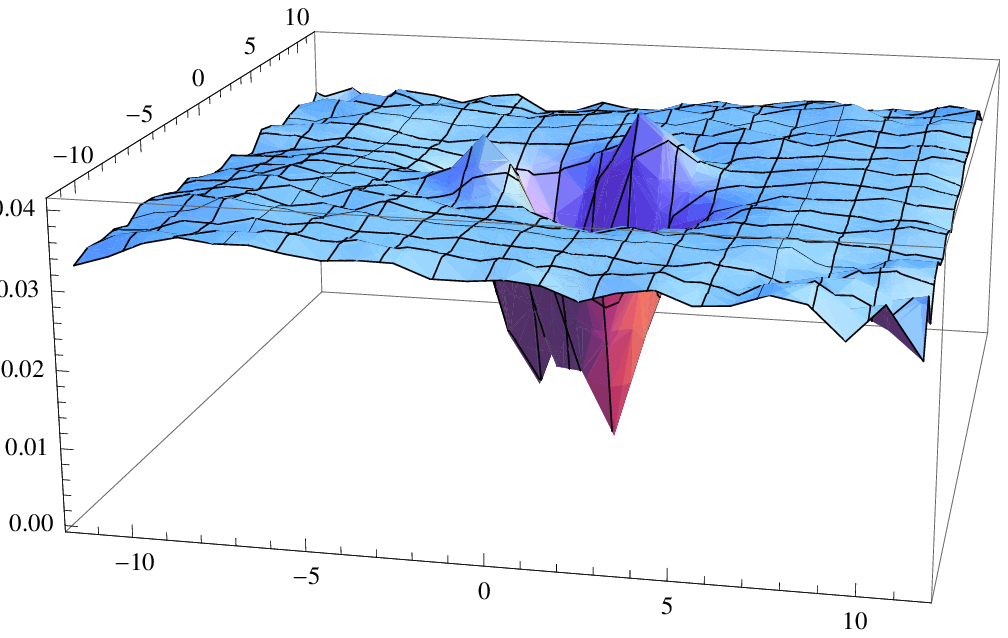}}
\caption{\label{viewd} Spatial variation of the vortex $d$ wave component, $\Delta_d(i)$, near the vortex core. 
$E_F=1.6$.}
\end{figure}
The superconducting term in the model (\ref{raghu}) is derived from the interaction term
\begin{eqnarray}
\hat V = \frac 1 4 \sum_{ij}\sum_{\nu,\nu'} V_{ij} 
\hat d^\dagger_{i\up, \nu} \hat d^\dagger_{j\dn, \nu} 
\hat d_{j\dn, \nu'} \hat d_{i\up, \nu'}\,,
\end{eqnarray}
and the gap equation is
\begin{equation}
\Delta_{ij}=\sum_{\nu}  V_{ij} \la \hat d_{j\dn, \nu} \hat d_{i\up, \nu} \ra\,. \label{gapequation}
\end{equation}
We here assume $V_{ij}$ to take on a {\it negative} value, $V$, for second neighbours 
and be equal to zero otherwise.
 The field operators are related to the excitation operators, $\hat\gamma_n$, with energy $E_n$, through
the equations \cite{deGennesbook} 
\begin{eqnarray}
\hat d_{i\up, \nu} &=& \sum_n \left[ u_{n,\nu}(i)\hat \gamma_n  - v^*_{n,\nu}(i) \hat \gamma^\dagger_n  \right] \\
\hat d_{i\dn, \nu} &=& \sum_n \left[ u_{n,\nu}(i)\hat \gamma_n  + v^*_{n,\nu}(i)\hat \gamma^\dagger_n  \right]\,. 
\end{eqnarray}  
The Bogolubov-de Gennes equations for the amplitudes $u$ and $v$, using the model  (\ref{raghu}), are:
\begin{eqnarray}
(E_n+E_F) u_{n,\nu}(i) &=& \sum_j \left[  - t_{ij\sigma, \nu} u_{n,\nu}(j) 
- t_4 \Theta_{ij} u_{n,\bar\nu}(j) 
+ \frac 1 4 \Delta_{ij} v_{n,\nu}(j) \right] \,,\label{Bdg1}\\
(E_n-E_F) v_{n,\nu}(i) &=& \sum_j \left[  \frac 1 4 \Delta^*_{ij} u_{n\nu}(j)  + t_{ij\sigma, \nu} u_{n\nu}(j) 
+ t_4 \Theta_{ij} u_{\bar\nu}(j) \right] \label{Bdg2}\,.
\end{eqnarray}  

Equations (\ref{Bdg1})-(\ref{Bdg2}) and  
(\ref{gapequation}) have been solved self-consistently, 
by iterations until convergence was achieved,  
in a 30$\times$30 lattice, with open boundary conditions.
The vortex core is located at the centre of a plaquette (see Figure \ref{esquema}).  
We take $V=-4$ and  
the chemical potential $E_F$ in the range 1.35-1.8, below.
The gap function describing a vortex can be written in the form \cite{franztesa}
\begin{equation}
\Delta_{ij} = \tilde\Delta_{\vec{ij}}(i)\  e^{i\theta_i}\,,
\end{equation}
where $\theta_i$ is the azimuthal angle of site $i$. We define the $s$ wave component
of $\Delta_{ij}$ as
\begin{equation}
\Delta_s(i)=
\tilde\Delta_{ \hat \vx + \hat \vy}(i) +
\tilde\Delta_{- \hat \vx - \hat \vy}(i) +
\tilde\Delta_{ \hat \vx - \hat \vy}(i) +
\tilde\Delta_{- \hat \vx + \hat \vy}(i) \,,
\end{equation}
and the $d$ wave component 
of $\Delta_{ij}$ as
\begin{equation}
\Delta_d(i)=
\tilde\Delta_{ \hat \vx + \hat \vy}(i)+
\tilde\Delta_{- \hat \vx - \hat \vy}(i) -
\tilde\Delta_{ \hat \vx - \hat \vy}(i) -
\tilde\Delta_{- \hat \vx + \hat \vy}(i) \,.
\end{equation}

\begin{figure}[t]
\vspace{1.3cm}
\centerline{\includegraphics[width=6.5cm]{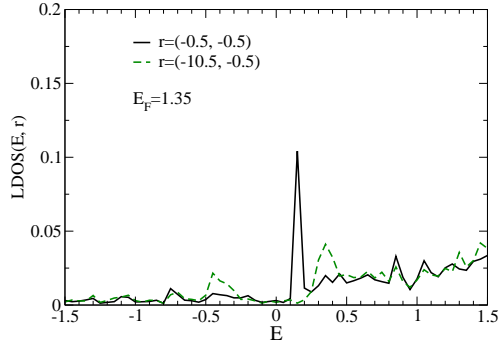}}
\caption{\label{ldose160} The LDOS as function of energy near the cortex core  [the point $(-0.5, -0.5)$] and
outside the vortex core [the point $(-10.5, -0.5))$].  The chemical potential $E_F=1.35$.} 
\end{figure}
\begin{figure}
\vspace{1.3cm}
\centerline{\includegraphics[width=6.5cm]{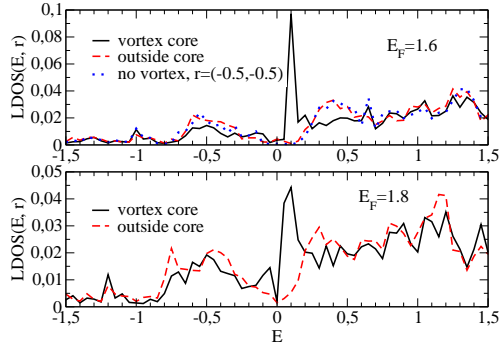}}
\caption{\label{ldose180} The LDOS as function of energy near the cortex core  [the point $(-0.5, -0.5)$] and
outside the vortex core [the point $(-10.5, -0.5))$].  The chemical potentials are $E_F=1.6$ (top) and  $E_F=1.8$ (bottom). The LDOS without vortex is also shown for comparison (top).  }
\end{figure}

\subsection{Results}

For this choice of $V_{ij}$ both components are always present, even 
in the uniform (no vortex) case. The $d$ wave component is, nevertheless, rather small, as can be seen
from
Figures \ref{views} and \ref{viewd}, which   show  $\Delta_s(i)$ and  $\Delta_d(i)$ in a region including 
 the vortex core.
The $d$ wave component displays  the  more complex behaviour:
 $\Delta_d(i)$ has four minima, symmetric with respect to rotations through the angle  $\pi$ (not $\pi/2$) around the 
vortex centre.  The $s$ wave component, on the other hand,  has  a single minimum.

\begin{figure}[t]
\centerline{\includegraphics[width=6.5cm]{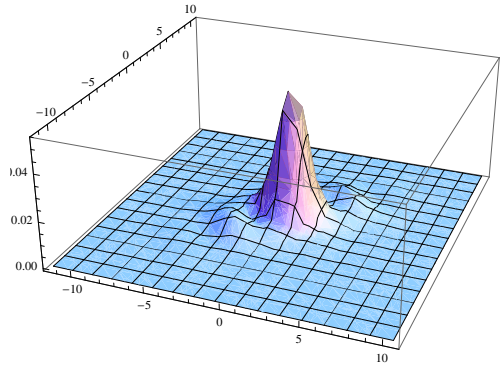}}
\caption{The spatial dependence of the function  $|u(\vr)|^2$ at the LDOS peak energy for
$E_F=1.6$. Wave functions close to the Fermi level are spatially  extended. } 
\label{excitation}
\end{figure}

Figures \ref{ldose160}-\ref{ldose180} show the local density of states, computed from 
equation (\ref{sempre}),  as a function of energy, 
at two different locations:  near the vortex centre [the point $(-0.5, -0.5)$] and 
outside the vortex core [the point $(-10.5, -0.5))$]. At the latter location, where $\Delta_{ij}$
is almost uniform, the LDOS displays the typical behaviour:
it is  small at low energy and has two maxima at energies close to $\pm \Delta$. Due to the 
band structure, the curve is not quite symmetrical around $E=0$. 
The LDOS at the vortex centre is not typical: contrary to that obtained in  the continuum model
above, it has a fairly sharp maximum at finite energy above the Fermi level. 
The peak energy is smaller than but of the order of $\Delta$ and is   
also dependent on the chemical potential.
In addition to this peak, the LDOS also displays a V-shaped background with weak 
oscillations which are due to finite size effects.   

The LDOS peak reveals the existence of localized 
excitations, at the vortex core,  at finite energy  and 
Figure \ref{excitation} shows one such localized excitation. 
The low-lying excitations are spatially  {\it delocalized}. 
Such behaviour can be qualitatively understood if we see the  core excitations
as Andreev bound states that form in the region of small $\Delta$ that is in contact
with  the region of large $\Delta$. The Andreev reflection problem has recently been
investigated in the context of the $s^\pm$ superconductivity \cite{us}. 
It was found that Andreev bound states form at finite energy
because of interference effects between waves in 
different bands of the superconductor.
In the continuum formulation used in the previous section,  
the bands were treated separately and such interference effects were lost.   
We note that 
the familiar Caroli-de Gennes-Matricon  expression $\Delta^2/E_F$ 
for the bound states energy cannot be straightforwardly used here because 
it applies to a  single electron parabolic band with the 
Fermi energy  measured from  the bottom of the band, assumed to be zero. 
The Fermi energy values, $E_F$, above, are not measured from the band bottom. 
The band structure has half-metal character, with hole and electron bands and 
quantum interference effects between them.

\begin{figure}[htb]
\vspace{1.3cm}
\centerline{\includegraphics[width=6.5cm]{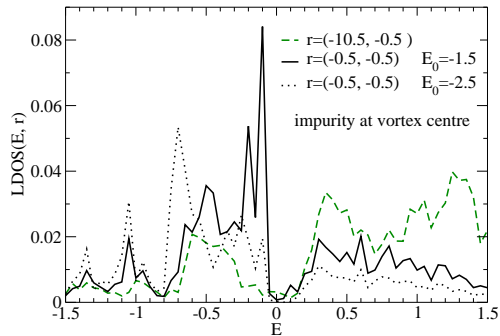}}
\caption{\label{atractiveimp} The LDOS versus energy at the 
(attractive) impurity site  $(-0.5, -0.5)$ 
near the vortex core,
and outside the vortex core [the point $(-10.5, -0.5))$].  The chemical potential $E_F=1.6$.
$E_0$ denotes the impurity potential. } 
\end{figure}

\begin{figure}[htb]
\vspace{1.3cm}
\centerline{\includegraphics[width=6.5cm]{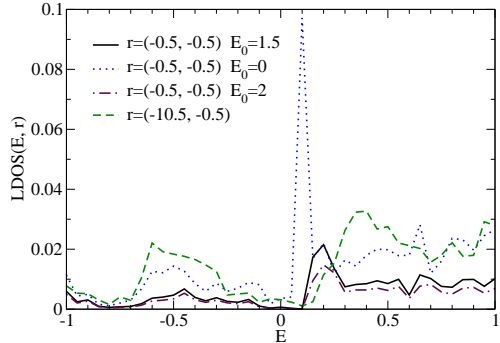}}
\caption{\label{repulsiveimp} The LDOS versus energy at the 
(repulsive)  impurity site  $(-0.5, -0.5)$  near the vortex core, 
and outside the vortex core [the point $(-10.5, -0.5))$].  The chemical potential $E_F=1.6$.
$E_0$ denotes the impurity potential. } 
\end{figure}

In the STM experiment of Ref \cite{experiment}, 
the vortex positions were not correlated with the positions of impurities on 
the sample surface.
We find that the presence of an impurity far from the vortex core only affects the 
LDOS in the impurity's vicinity. This is in agreement with results in other systems
where the effect of impurities is rather local \cite{fischer}. However, we may
consider a situation where the vortices are pinned by impurities even though
there seems to be bulk pinning in the experimental system studied in Ref. \cite{experiment}.
The effect of an impurity located at the vortex core is shown in Figures  \ref{atractiveimp}
 and  \ref{repulsiveimp}.
While an impurity located outside the vortex has little effect on the core LDOS peak, 
an attractive impurity close to the core shifts the peak to negative energy, 
as shown in Figure \ref{atractiveimp}, whereas the  LDOS outside the core remains unchanged.

A repulsive impurity close to the vortex core has  a stronger effect:
the former LDOS peak is almost suppressed and shifted to higher  energy, 
as shown in Figure \ref{repulsiveimp}, while the  LDOS outside the core remains unchanged.
The location of the low energy peak is therefore shifted depending on the type and
strength of the impurity and on the band filling. 

\section{Summary and discussion}

Both the continuum and the lattice formulations presented here lead to localized
excitations in the vortex core. In the first case there is a zero energy peak in
the LDOS at the vortex location resulting from the addition of two off-centre peaks, 
one from the electron band and one from the hole band.
In the lattice formulation, interference effects from the two bands
are taken into account, and only one gap function is introduced containing
both bands. The low energy states are extended
and  localized modes at the vortex core appear at higher energy producing a 
LDOS peak deviated from the Fermi level. 
This may be understood 
from the recent prediction that Andreev bound states have  finite
energy in the s$^\pm$ scenario. 
It also  implies that the density of (low lying
extended) states should be proportional to the density of vortices, hence
proportional to the applied magnetic field, $H$. 
Therefore, if the quasiparticle mean-free-path is longer than intervortex spacing
\cite{kubert} then linear behavior of zero temperature heat transport, 
$\kappa_0/T(H) \propto H$, may be expected \cite{taillefer}.

Our calculations 
shows that the effect of impurities at the vortex location has a significant
effect on the peak in the LDOS but does not remove it.

The prediction of localized excitations in the vortex core 
is in conflict with the  experimentally observed LDOS, with a STM, 
in  BaFe$_2$Co$_{0.2}$As$_2$, where no peak in $\rho(E,\vr)$ was observed at the vortex centre \cite{experiment}. 
Indeed, the measured differential conductance displays only the V-shaped background 
line, similar to that visible in figures \ref{ldose160} and \ref{ldose180} 
if one disregards the central peak and weak oscillations (due to finite size effects). 
The observed LDOS away from the vortex centre is similar to that in figures
\ref{ldose160} and  \ref{ldose180}, showing a  depression below the gap energy.  

In other superconductors either conventional or unconventional there are low energy
peaks in the LDOS at the vortex location. In the case of s-wave conventional superconductors
there are peaks corresponding to the Caroli-de Gennes-Matricon states \cite{Gygi,machida} and
in d-wave unconventional superconductors there are two low-energy peaks that replace the coherence
peaks at the vortex location and immediate vicinity \cite{fischer}. Most superconductors have
some sort of disorder leading in general to positional disorder of the vortices. This leads
to an increase of the low energy LDOS and possible closing of the gap, in the s-wave superconductors,  
and to a finite density of states in the d-wave case \cite{volovik,lages}. 
The effect of a moderate concentration
of impurities was considered in the d-wave case showing that the effect of the vortex
disorder is dominant \cite{lages2,vafek}. However, in the very dirty superconductor 
Nb$_{1-x}$Ta$_x$Se$_2$, 
where the quasiparticle mean path is smaller than the coherence length, the
low energy peaks in the LDOS have been observed to give way to an almost flat conductance 
similar to the normal state one
\cite{renner}. Similar results were obtained in the case
of MgB$_2$,  even though in this system the mean path is expected to be large
\cite{eskildsen}.

In the course of completion of this work, two preprints appeared where the LDOS in a vortex lattice 
was theoretically studied \cite{hu,jiang}. Our results for a single vortex are in agreement with those
studies.

\end{document}